# Origin of the density wave instability in trilayer nickelate $La_4Ni_3O_{10}$ revealed by optical and ultrafast spectroscopy

Shuxiang Xu [1*], Cui-Qun Chen [2*], Mengwu Huo[2], Deyuan Hu[2],
Hao Wang [1], Qiong Wu [1], Rongsheng Li [1], Dong Wu [3], Meng Wang [2], Dao-Xin Yao [2+],
Tao Dong [1+], Nanlin Wang [1, 3 +]

[1]*International Center for Quantum Materials, School of Physics, Peking University, Beijing 100871, China*

[2] *Guangdong Provincial Key Laboratory of Magnetoelectric Physics and Devices, School of Physics, Sun Yat-Sen University, Guangzhou, Guangdong 510275, China*

[3]*Beijing Academy of Quantum Information Sciences, Beijing 100193, China*

+Corresponding authors: yaodaox@mail.sysu.edu.cn, taodong@pku.edu.cn, nlwang@pku.edu.cn

* These two authors contributed equally to this work.

# Abstract

In the intricate phase diagram of unconventional superconductors characterized by intertwined electronic orders and superconductivity, a key step in understanding the superconducting mechanism is to investigate the parent compounds from which superconductivity emerges through doping or pressure. In this study, we employed optical spectroscopy and ultrafast reflectivity measurements to examine the density wave instability in the trilayer nickelate $La_4Ni_3O_{10}$, which displays pressure-induced superconductivity up to 30 K. Our optical spectroscopy measurements reveal that $La_4Ni_3O_{10}$ behaves as a metal with a high plasma frequency. Upon cooling, we observed a distinct formation of a density wave energy gap in both optical conductivity and pump-probe measurements. The gap feature is more pronounced compared to the bilayer nickelate $La_3Ni_2O_7$. Through a comparison of the experimentally determined plasma frequency with first-principles calculations, we classify $La_4Ni_3O_{10}$ as a moderately electron-correlated material, resembling the parent compound of iron-based superconductors but exhibiting weaker correlation than the bilayer nickelate $La_3Ni_2O_7$. The enhanced gap feature and weaker electronic correlation in $La_4Ni_3O_{10}$ may explain its lower superconductivity transition temperature under high pressure. These findings significantly advance our comprehension of the density wave and superconductivity mechanisms in the trilayer nickelate $La_4Ni_3O_{10}$.

## Introduction

In numerous unconventional superconductors, such as cuprates and iron-based superconductors, the presence of charge and/or spin orders is intimately connected to their superconducting properties[1-6]. Understanding the nature of the charge/spin orders and how superconductivity emerges from the background of the spin/charge orders is a key ingredient in understanding the superconductivity mechanism. Recently, the discovery of high-temperature superconductivity in nickelate has attracted tremendous interest in the condensed matter community[7-24]. So far, the discovery of nickelate superconductors is extremely rare. The first nickelate superconductor discovered is the thin films of infinite-layer nickelates $Nd_{1-x}Sr_xNiO_2$ with $T_c$ ~ 9-15 K[12, 13, 16]. After that, high temperature superconductivity in $La_3Ni_2O_7$ and $La_4Ni_3O_{10}$ was observed under pressure by many experimental teams[7, 9-11, 19, 21, 25]. The maximal superconducting transition temperature $T_c$ reaches 80 K in $La_3Ni_2O_7$, but only 30 K in $La_4Ni_3O_{10}$ crystal. The common feature among the three is that all of them belong to the Ruddlesden-Popper ($La/Nd_{n+1}Ni_nO_{3n+1}$) compound family with $n$ = 2, 3 and ∞, respectively. It is worth mentioning that the superconducting transition temperature in $La_3Ni_2O_7$ and $La_4Ni_3O_{10}$ is comparable to that in cuprate superconductors[26-28]. However, in multi-layer cuprate superconductors, $T_c$ increases with the number of $CuO_2$ layers up to three. This difference indicates that the superconducting mechanism in nickelates is distinct from that in cuprate superconductors. How to understand the mechanism of high temperature superconductivity in nickelates and lower $T_c$ in $La_4Ni_3O_{10}$ than $La_3Ni_2O_7$ remains an open question[14, 29-43]. Similar to cuprates and iron-based superconductors, $La_3Ni_2O_7$ and $La_4Ni_3O_{10}$ also display complex competing orders, including density wave and superconductivity[2, 4, 8, 44, 45]. Therefore, understanding the density wave instability in $La_3Ni_2O_7$ and $La_4Ni_3O_{10}$ is a crucial step toward comprehending the mechanism of superconductivity. In $La_3Ni_2O_7$, the electron-phonon coupling is reported to be weak, but electronic correlation is extremely strong, which may be responsible for the formation of density wave and superconductivity[15, 29, 46]. In $La_4Ni_3O_{10}$, the intertwined charge density wave (CDW) and spin density wave (SDW) coexist, and density functional theory shows that susceptibility reaches maxima near the CDW/SDW wave vector, indicating Fermi surface nesting (FSN) may be the main mechanism of CDW/SDW[45]. However, the optical experimental evidence about $La_4Ni_3O_{10}$ remains rare. Whether the mechanism is from FSN or not requires further experimental investigation.

Optical spectroscopy, a technique sensitive to probing the electronic properties of bulk materials, plays a crucial role in revealing the essence of the electronic structure and CDW/SDW transition[47, 48]. On the other hand, the ultrafast pump-probe technique is commonly used to probe the relaxation process of excited quasiparticles, which is believed to be determined by the same interactions that govern the equilibrium properties and are also sensitive to the gap opening near the Fermi level. Ultrafast spectroscopy has found wide application in investigating the density wave states, nematic states, superconducting states, and beyond[49-53].

In this work, we present a comprehensive study on $La_4Ni_3O_{10}$ crystal, employing a combination of optical spectroscopy and ultrafast pump-probe measurements. Our optical spectroscopy shows that $La_4Ni_3O_{10}$ crystal is metallic with a high plasma frequency > 2 eV. We observe the clear formation of an energy gap when the system enters into the CDW/SDW state. The energy gap is estimated to be 61 meV by the peak position from the density wave component. Additionally, a large Drude component and carrier scattering rate drop sharply near the phase transition, similar to the phenomenon of $BaFe_2As_2$ [54], indicating the Fermi surface nesting may be the driven force of the CDW/SDW state. Combining experimental results with the first-principles calculation, we obtained the kinetic ratio $K_{\mathrm{exp}}/K_{\mathrm{band}} \sim 0.29$, which reveals $La_4Ni_3O_{10}$ belongs to a moderately electronic correlation material resembling the parent compound of iron-based superconductors[15, 55]. Our ultrafast pump-probe experiment has shown that the amplitude reduces rapidly to zero, and the relaxation time diverges around the CDW/SDW transition temperature, consistent with the typical characteristics of a second-order phase transition. Utilizing the Rothwarf-Taylor (RT) model [56-58], the energy gap has been suggested to be 55 meV, close to the result of optical spectroscopy. Our study, provides key information for understanding the nature of CDW/SDW state and unconventional superconductivity in $La_4Ni_3O_{10}$ crystal.

# Experimental methods

High-quality single crystals of $La_4Ni_3O_{10}$ were synthesized by high pressure floating zone method with a fixed oxygen pressure of 20 bar. The resistivity measurement was performed in a Quantum Design physical property measurement system (PPMS) using a standard four-contact method with electronic current parallel to the *ab* plane. Optical reflectance measurements of $La_4Ni_3O_{10}$ were performed on the Fourier transform infrared spectrometer Bruker 80V in the frequency range from 40 to 30 000 $cm^{-1}$. The value of reflectivity $R(\omega)$ was obtained by an in situ gold and aluminum evaporation technique. The real part of conductivity $\sigma_1(\omega)$ is calculated by the Kramers-Kronig transformation of $R(\omega)$. A Hagen-Rubens relation is employed for low-frequency extrapolation and an x-ray atomic scattering factor is used for high-frequency extrapolation[59]. Time-resolved reflectivity experiments were conducted using an optical fiber oscillator with a center wavelength of 1560 nm, a repetition rate of 80 MHz, and a pulse duration of 120 fs. The pump and probe wavelengths are 780 nm and 1560 nm, respectively. Both of them are vertically polarized. The spot size of the pump and probe beams are focused to 120 μm and 80 μm on the sample, respectively. The vertically polarized pump beam was chopped at 733 kHz by an acousto-optic modulator to facilitate lock-in detection. The pump fluence was tuned to 0.2 $\mu J/cm^2$ while the probe fluence was reduced to only 10% of the pump. The weak reflectivity signal was detected by an amplified detector.

Density functional theory (DFT) calculations were performed using the Vienna *ab initio* Simulation Package (VASP)[60] with local density approximation (LDA) exchange correlation potential[61, 62]. We adopted the projector augmented wave with a plane-wave cutoff energy of 600 eV. The experimentally measured lattice constants at ambient pressure[8] were used in our calculations and the positions of all atoms were fully relaxed until the force on each atom was less than 0.001 eV/Å. The energy convergence criterion was set at $10^{-7}$ eV for the electronic self-consistent loop and a Γ-centered 20×20×19 k-mesh grid was employed. The broadening factor was set to be 120 $cm^{-1}$ in computing $\sigma_1(\omega)$, corresponding to the experimental Drude width at 15 K. The interband optical conductivity was calculated based on Kubo-Greenwood formula as implemented in WANNIER90 package[63-65] and the calculation was performed on a 500×500×200 k-mesh. The plasma frequency was calculated by[66]

$$\omega_{p(\alpha\beta)}^2 = -\frac{4\pi e^2}{V\hbar^2}\sum_{n,k} 2g_{\boldsymbol{k}}\frac{\partial f(\epsilon_{n\boldsymbol{k}})}{\partial\epsilon}\left(\boldsymbol{e}_\alpha\frac{\partial\epsilon_{n\boldsymbol{k}}}{\partial\boldsymbol{k}}\right)\left(\boldsymbol{e}_\beta\frac{\partial\epsilon_{n\boldsymbol{k}}}{\partial\boldsymbol{k}}\right),$$

where, $g_{\boldsymbol{k}}$ are the weight factors of the **k** points, $f(\epsilon_{n\boldsymbol{k}})$ is Fermi-Dirac occupation function. Since the optical conductivity was measured in *ab* plane, the calculated optical conductivity is given by the average of $\sigma_{xx}(\omega)$ and $\sigma_{yy}(\omega)$.

# Results and Discussions

Figure 1(a) shows the electronic transport of $La_4Ni_3O_{10}$ single crystal. As temperature drops, the resistivity value decreases, which is consistent with the metallic behavior. When the temperature reduces to 140 K, an abnormal hump appears in the resistivity curve. According to the previous report, this abnormal hump originates from the intertwined charge and spin density wave (DW)[45]. By differentiating the resistivity, the transition temperature of DW $T_{DW}$ was obtained to be approximately 136 K, which is close to the results reported previously. The inset of Fig. 1(a) shows the crystal structure of $La_4Ni_3O_{10}$. As we can see, in a unit cell, three Ni-O layers form a trilayer structure. In each Ni-O layer, the Ni atom with the nearest six oxygen atoms forms the $NiO_6$ octahedron. These $NiO_6$ octahedrons connect to each other by sharing the apical oxygen atoms. Figure 1(b) displays the reflectivity $R(\omega)$ of $La_4Ni_3O_{10}$ below 5000 $cm^{-1}$ at several selected temperatures. For all temperatures, with frequency decreasing from 5000 $cm^{-1}$, the reflectivity increases gradually and approaches unity in the zero-frequency limit. For $T \geq 150$ K, the reflectivity below 5000 $cm^{-1}$ grows up gradually with temperature decreasing from 300 K to 150 K. For $T \leq 120$ K, $R(\omega)$ shows a substantial suppression from 500 to 3000 $cm^{-1}$, which is a strong evidence for the formation of the energy gap. At low frequency near 500 $cm^{-1}$, the $R(\omega)$ increases faster than that above 150 K, which produces a sharp plasma edge. This characteristic implies that the Fermi surfaces are only partially gapped and $La_4Ni_3O_{10}$ maintains metal properties below DW transition. In addition, at about 600 $cm^{-1}$ (74 meV), $R(\omega)$ shows a dip in all temperatures. This dip sharpens gradually with temperature decreases, which is consistent with the feature of phonon. The inset of Fig. 1(b) shows the reflectivity over a large energy scale of 40-30000 $cm^{-1}$. As we can see, reflectivity reduces monotonically to about 0.25 when frequency increases to 15000 $cm^{-1}$. Here, we focus on the change of $R(\omega)$ at low frequency. Figure 1(c) shows the real part of optical conductivity spectra $\sigma_1(\omega)$ below 3000 $cm^{-1}$. The Drude-like conductivity can be observed at all temperatures. For $T$ = 120 K (< $T_{DW}$ ~ 136 K), the Drude component is obviously suppressed at low frequency, and a weak peak appears near 900 cm$^{-1}$. For $T$ = 80 K and 15 K, the Drude components are severely suppressed at low frequency, and a Lorentz-like peak at about 1000 cm$^{-1}$ can be seen clearly. These spectral features are very similar to the optical response of Ba/SrFe$_2$As$_2$ crystal [67].
To qualitatively analyze the variation of the free-carrier component and the density wave gap with temperature, we fit the optical conductivity $\sigma_1(\omega)$ by the Drude-Lorentz model for all temperatures:

$$\sigma_1(\omega) = \sum_i \frac{\omega_p^2}{4\pi} \frac{\frac{1}{\tau_i}}{\omega^2 + \left(\frac{1}{\tau_i}\right)^2} + \sum_j \frac{S_j^2}{4\pi} \frac{\frac{1}{\tau_j}\omega^2}{(\omega_j^2 - \omega^2)^2 + \omega^2\left(\frac{1}{\tau_j}\right)^2}$$

Here, the first term is the Drude component, and the second is the Lorentz component. Notably, $\sigma_1(\omega)$ shows a noticeable kink at about 350 cm$^{-1}$ for all temperatures, revealing a Lorentz component possibly contributing to it. Apart from that, a small sharp peak along with a very wide width is observed clearly at 600 cm$^{-1}$, which is considered to result from the overlay of a sharp phonon and a Lorentz component. For $T \geq$ 150 K, we use two Drude components for the free-carrier response due to the multiband characteristic of La$_4$Ni$_3$O$_{10}$. Three Lorentz components were used to fit the anomaly envelope at low frequency ($\omega$ < 1000 cm$^{-1}$). For $T \leq$ 120 K, we also use two Drude components for the free-carrier response and three Lorentz components for low frequency. Besides that, a Lorentz component was added to fit the newly emerging DW peak near 1000 cm$^{-1}$. According to previous reports, the energy gap could be estimated by the central frequency of the Lorentz model [54, 67]. The inset of Figure 1(c) shows the fitting result of $\sigma_1(\omega)$ in $T$ = 15 K. For La$_4$Ni$_3$O$_{10}$, the central frequency of DW response is about 990 cm$^{-1}$ at 15 K. Thus, the energy gap of La$_4$Ni$_3$O$_{10}$ obtained by infrared spectroscopy is about 61 meV which is nearly 3 times as large as ARPES experiment [68]. Then, we obtain the ratio of $2\Delta/k_BT_{DW}$ = 10.22, much larger than the weak-coupling BCS value of 3.52 but close to the value in La$_3$Ni$_2$O$_7$ 10.14 [15]. High-frequency $\sigma_1(\omega)$ was fitted by the same Lorentz component for all temperatures. The experimental $\sigma_1(\omega)$ could be well fitted for all temperatures, as shown in the inset of Fig. 1(c) and Fig. 5. The detailed fitting parameters at low frequency are summarized in Table. 1 and Table. 2. For the two Drude models, the overall plasma frequency could be considered as coming from the two different channels and calculated by $\omega_p = \sqrt{\omega_{p1}^2 + \omega_{p2}^2}$. The overall plasma frequency we obtained was $\omega_p \approx$ 16860 cm$^{-1}$ at 300 K reduced to 10880 cm$^{-1}$ at 15 K. Fig. 1(f) shows the temperature evolution of $\omega_p^2$ and scattering rate $1/\tau$ for two Drude components. Both of them are normalized to their 300 K values. It is known that the Drude spectral weight $\omega_p^2$ is proportional to $n/m_{eff}$, where $n$ is the carrier density and $m_{eff}$ is the effective mass, respectively. Provided that the $m_{eff}$ does not change with temperature, it means that the carrier density (Fermi surface) from the Drude2 component is reduced (removed) by about 97% after the DW transition while Drude1 component changes slightly. In the meantime, the scattering rate from the Drude1 and Drude2 component is decreased by about 53% and 75%, respectively. The reduction of scattering rate leads to a metallic state in the gapped DW state.

To avoid the fitting error from the Drude-Lorentz model, the spectral weight $S(\omega)$ was calculated by integrating the conductivity $\sigma_1(\omega)$ ($S(\omega) = \int_0^{\omega} \sigma_1(\omega) d\omega$) as shown in Fig. 1(d). In the zero-frequency limit, the spectral weight for $T \leq 120$ K is higher than that for $T \geq 150$ K, implying the Drude components become narrow and high below $T_{SDW}$. Between 500 and 3000 cm$^{-1}$, the spectral weight for $T \leq 120$ K (purple curve) is obviously lower than that at $T = 150$ K (green curve) which means the opening of energy gap. To further estimate the size of the energy gap, frequency-dependent spectral weight ratio of $S$(15 K)/$S$(150 K) is shown in the inset of Fig. 1(d). In the zero-frequency limit, the huge $S$(15 K)/$S$(150 K) value is correlated to the narrow Drude component at low temperatures. With increasing frequency, the value of $S$(15 K)/$S$(150 K) reaches a minimum of about 800 cm$^{-1}$, which is attributed to the spectral weight loss at low temperatures. It further evidences that the DW phase transition is accompanied by the opening of an energy gap on the Fermi surface. And the energy gap size $2\Delta$ is close to 800 cm$^{-1}$. A similar phenomenon has been observed in some parents of Fe-based superconductors [54]. Due to the spectral weight being mainly from the Drude component, the feature of spectral weight loss is consistent with the suppression of $\omega_p^2$ and $1/\tau$ at low temperatures. With further increasing frequency, the ratio of $S$(15 K)/$S$(150 K) increases monotonically and approaches unity at 6000 cm$^{-1}$, revealing that the spectral weight is gradually recovered over a broad frequency range. Figure 1(e) shows the $S(T)/S$(300 K) at several selected cutoff frequencies as a function of temperature. For $\omega_c = 600$ cm$^{-1}$ and 1000 cm$^{-1}$, $S(T)/S$(300 K) increases upon cooling from 300 K to 150 K, mainly originating from the narrowing of the Drude response. Below $T_{DW}$, $S(T)/S$(300 K) decreases due to gap opening consistent with the suppression of Drude spectral weight. For $\omega_c = 9000$ cm$^{-1}$, $S(T)/S$(300 K) is almost temperature-independent, indicating the spectral weight lost at low frequency is recovered at 9000 cm$^{-1}$.

For $La_3Ni_2O_7$, the electronic correlation is very strong and significant for the formation of charge density wave and superconductivity. Therefore, it is indispensable to estimate the strength of electronic correlation in $La_4Ni_3O_{10}$. It is known that the strength of electronic correlation is inversely proportional to the ratio $K_{exp}/K_{band}$, where $K_{exp}$ and $K_{band}$ refer to the experimental kinetic energy and theoretical kinetic energy from band theory, respectively. Then, we performed the first-principles density functional theory (DFT). The electronic band structure in Fig. 2(b) shows that multiple bands cross the Fermi level, which is more complicated than that in $La_3Ni_2O_7$. The calculated $\sigma_1(\omega)$ indicates that there exists low-energy interband transitions in $La_4Ni_3O_{10}$, as shown in Fig. 2(c) and its inset. Thus, the intraband and interband transitions are overlapped with each other at low frequency. According to the previous report [15], the $K_{exp}/K_{band}$ is proportional to $\omega^2_{p,exp}/\omega^2_{p,cal}$. The total Drude component gives $\omega_{p,exp} = 16400$ cm$^{-1}$ (2.03 eV) at $T = 150$ K, whereas DFT calculations yield $\omega_{p,cal} = 30200$ cm$^{-1}$ (3.74 eV). Consequently, we obtain the ratio $K_{exp}/K_{band} \sim 0.29$, which is one order of magnitude larger than that in $La_3Ni_2O_7$ (0.022). This result reveals that the electronic correlation of $La_4Ni_3O_{10}$ is weaker than that of $La_3Ni_2O_7$. In addition, we summarize $K_{exp}/K_{band}$ for $La_4Ni_3O_{10}$ and various other materials shown in Fig. 2(d)[15, 55]. As we can see, $K_{exp}/K_{band}$ in $La_4Ni_3O_{10}$ lies between conventional metals and mott insulators, closely similar to the parent compounds of iron-based superconductors such as $BaFe_2As_2$. The above result indicates $La_4Ni_3O_{10}$ belongs to a moderately correlated material. In fact, previous theoretical calculations have reported that the correlation in $La_{n+1}Ni_nO_{3n+1}$ is layer-dependent[38]. The recent ARPES measurement further suggests that the electronic correlation strength is orbital-dependent in $La_3Ni_2O_7$[29]. The inner $NiO_2$ layer in $La_4Ni_3O_{10}$ is weakly correlated due to more hole doping, where the Ni cations have a higher valence. However, the outer $NiO_2$ layers are more correlated, resembling the $NiO_2$ layer in $La_3Ni_2O_7$. Our experimental result is consistent with theoretical analysis. Electronic correlation significantly impacts physical properties such as DW and superconductivity. The weak electronic correlation in $La_4Ni_3O_{10}$ compared with $La_3Ni_2O_7$ might be the cause of the relatively lower $T_c$ in $La_4Ni_3O_{10}$ under pressure. In addition, the previous theoretical study reported that the $K_{DMFT}/K_{DFT}$ of infinite compound $LaNiO_2$ (0.5~0.6) is larger than that in $La_4Ni_3O_{10}$ (0.29) further revealing that $T_c$ is positively related to the strength of electronic correlation [69, 70].

To get more information about the DW, we performed the ultrafast pump-probe experiment on $La_4Ni_3O_{10}$. Figure 3(a) shows the time-dependent $\Delta R/R = [R(t)-R(0)]/R(0)$ along the *ab* plane at different temperatures, where $R(0)$ is the reflectivity measured before the pump pulse arrives. At low temperatures, when the pump light reaches the sample, the amplitude of $\Delta R/R$ increases quickly and then drops back to the equilibrium state within several picoseconds. With increasing temperature to 136 K, the relaxation time of excited quasiparticles back to the equilibrium state reaches its maximum. When the temperature increases to 150 K, the $\Delta R/R$ completely changes its sign from negative to positive. The extracted transition temperature of 136 K agrees with the value of $T_{DW}$ obtained from electrical transport. With further increasing temperature, $\Delta R/R$ keeps positive until room temperature. The temperature evolution of $\Delta R/R$ can be seen clearly in the contour map Fig. 3(b). From Fig. 3(b), we know that $\Delta R/R$ reaches a maximum at near 132 K, corresponding to the anomalous hump in the resistivity curve in Fig. 1(a). Figure 3(c) shows the typical transient variation in $\Delta R/R$ below 150 K. Here, the single exponential function can't fit the data well, thus we choose a double-exponential function to fit $\Delta R/R$. For $T <$ 150 K, the relaxation process is well fitted with a double-exponential function convoluted with a Gaussian laser pulse,

$$\frac{\Delta R}{R}(t) = \frac{1}{\sqrt{2\pi\omega}} exp(-\frac{t^2}{2\omega^2}) \otimes [A_1\, exp\,(-t/\tau_1) + A_2\, exp\,(-t/\tau_2) + C]$$

where $A_1$ ($A_2$) and $\tau_1$ ($\tau_2$) refer to the amplitude and relaxation time of the fast (slow) decay process. $\omega$ is the laser pulse temporal duration, C is a constant offset. The solid line in Fig. 3(c) is the fitted result utilizing this above formula which coincides well with the experimental data. The fitting parameters of $A_i$ and $\tau_i$ ($i$ = 1 and 2) at different temperatures are summarized in Figs. 4(a)-(d). According to earlier studies, the fast component is primarily associated with strongly coupled phonons scattering, whereas the slow component corresponds to the coupling of the rest of lattice [53, 57]. For the fast process, the amplitude $A_1$ in Fig. 4(a) decreases rapidly near $T_{DW}$. When the temperature is close to $T_{DW}$, the DW gap decreases and thermally excited quasiparticle increases in number hindering the photoexcitation of ground-state quasiparticle. In addition, $\tau_1$ in Fig. 4(b) is almost a constant at low temperature but increases sharply near $T_{DW}$ and then drops again above $T_{DW}$. The divergence of $\tau_1$ near $T_{DW}$ is the typical phonon bottleneck effect [56]. The rapid decrease of $A_1$ and divergence of $\tau_1$ appears simultaneously near the same temperature $T_{DW}$ indicating $La_4Ni_3O_{10}$ has an energy gap, and its gap is gradually closed near the critical temperature. A similar anomalous phenomenon has also been observed in $CsV_3Sb_5$, which exhibits an unconventional CDW gap [71]. To extract the gap value after the DW transition, we analyze the fast process $A_1$ and $\tau_1$ by the RT model, which has been widely applied in superconductors and density wave materials, where the formation of a gap in the density of states results in a relaxation bottleneck of the photoexcited quasiparticles[56-58]. The RT model is given by following equations[57, 71, 72]:

$$\tau(T) \propto \frac{\ln\left[g + e^{-\Delta(T)/k_B T}\right]}{\Delta(T)^2},$$

and

$$A(T) \propto \frac{\Phi \big/ [\Delta(T) + k_B T/2\,]}{1 + \Gamma\sqrt{2k_B T/\pi\Delta(T)}\,e^{-\Delta(T)/k_B T}},$$

where a BCS-like gap is assumed and gives $\Delta(T) = \Delta_0\sqrt{1 - T/T_{DW}}$, $\Phi$ is the pump fluence, $g$ and $\Gamma$ are the fitting parameters. The red solid lines in Figs. 4(a) and (b) are the fitting results. It gives an energy gap of $\Delta_0 \approx 55$ meV, which is close to the value of 61 meV obtained from infrared spectroscopy. The energy gap in $La_4Ni_3O_{10}$ is smaller than the value of the SDW gap (~ 70 meV) in $La_3Ni_2O_7$ obtained by ultrafast pump-probe measurement[73].

Finally, we discuss several implications of our observations. The first one is the difference of density wave in $La_3Ni_2O_7$ and $La_4Ni_3O_{10}$. Both charge density wave and spin density wave exist in the two compounds. However, for $La_3Ni_2O_7$, $T_{CDW}$ is lower, nearly 40 K, than $T_{SDW}$. Thus, SDW or CDW could be distinguished easily from the phase transition temperature. For $La_4Ni_3O_{10}$, the SDW and CDW develop simultaneously, which indicates charge order is strongly entangled with spin density wave. For typical CDW materials, the ultrafast spectroscopy always displays an amplitude mode which appears below $T_{CDW}$[50, 74]. Here, we did not observe any signature of amplitude mode in $La_4Ni_3O_{10}$. This means that the amplitude mode of CDW may be weak in $La_4Ni_3O_{10}$. For $La_3Ni_2O_7$, the optical spectroscopy reveals the opening of an energy gap with $\Delta$ ~ 50 meV only below $T_{CDW}$ rather than $T_{SDW}$[15]. The ultrafast study by Li *et al.* shows a weak kink below $T_{CDW}$[46], but Meng *et al.* reported that the relaxation time exists a strong divergence near $T_{SDW}$[73]. It indicates that $La_3Ni_2O_7$ crystal is sample-dependent. In addition, electronic transport experiments show that the amplitude of the kink near $T_{DW}$ in $La_3Ni_2O_7$ is significantly weaker than that in $La_4Ni_3O_{10}$[44, 45]. Pump-probe measurement reveals a clear diverging behavior around $T_{DW}$, indicating a gap opening in the density of states, which is absent in $La_3Ni_2O_7$ [46]. Furthermore, our optical conductivity measurements provide evidence of an absorption peak around 120 meV induced by the density wave gap opening, which is not observed in $La_3Ni_2O_7$ [15]. These observations suggest that the density wave gap in $La_4Ni_3O_{10}$ is more pronounced. For $La_4Ni_3O_{10}$, both optical spectroscopy and ultrafast measurements consistently display the opening of energy gap with $\Delta \sim 55$ meV below $T_{DW}$. Although the size of DW gap in $La_4Ni_3O_{10}$ is close to the CDW gap in $La_3Ni_2O_7$, the strength of electronic correlation in $La_4Ni_3O_{10}$ is smaller, nearly an order of amplitude, than $La_3Ni_2O_7$. The more hole doping in the inner $NiO_2$ layer of $La_4Ni_3O_{10}$ is responsible for the reduced electronic correlation. The second issue is the huge similarity of density wave and energy scales between $La_4Ni_3O_{10}$ and parent compound $BaFe_2As_2$ of iron-based superconductors. On the one hand, $T_{DW}$ (~ 136 K) in $La_4Ni_3O_{10}$ is nearly equal to that (~ 138 K) in $BaFe_2As_2$. And the size of energy gap is nearly the same [54, 75]. Furthermore, both $La_4Ni_3O_{10}$ and $BaFe_2As_2$ belong to the moderately correlated materials. On the other hand, when applying pressure, both $La_4Ni_3O_{10}$ and $BaFe_2As_2$ display superconductivity with nearly the same maximum $T_c \sim 30$ K[76, 77]. The third one is the origin of density wave instability in $La_4Ni_3O_{10}$. We observed that the formation of DW in $La_4Ni_3O_{10}$ is accompanied by the sharp reduction of the carrier density and the scattering rate similar to the phenomenon of $BaFe_2As_2$ indicating the Fermi surface nesting possibly be the driven force of DW[54]. Meanwhile, DFT result shows that susceptibility has maxima near DW vectors[45]. Therefore, Fermi surface nesting picture responsible for the formation of DW is favored.

# Conclusion

In summary, we comprehensively studied the static optical spectroscopy and ultrafast dynamics on $La_4Ni_3O_{10}$ crystal. Our work reveals that the Fermi surface opens an energy gap ($\Delta \sim 61$ meV) near the $T_{DW}$. The formation of DW is accompanied by the sharp reduction of the Drude spectral weight and scattering rate, indicating that the Fermi surface nesting may be the origin of the DW. In addition, the ultrafast dynamic measurement gives $\Delta \sim 55$ meV, consistent with the result of optical spectroscopy. Moreover, $La_4Ni_3O_{10}$ is properly classified as being in the moderate correlation regime, which is less correlated than $La_3Ni_2O_7$. The decrease of electronic correlation in $La_4Ni_3O_{10}$ possibly explains its lower superconducting $T_c$ under pressure.

# Acknowledgments

This work was supported by National Natural Science Foundation of China (Grant Nos. 12488201, 12174454, 92165204), the National Key Research and Development Program of China (Grant Nos. 2022YFA1403901, 2021YFA1400201, 2023YFA1406500 and 2022YFA1402802), the Guangdong Basic and Applied Basic Research Funds (Grant Nos. 2024B1515020040, 2021B1515120015), Guangzhou

Basic and Applied Basic Research Funds (Grant No. 2024A04J6417), Guangdong Provincial Key Laboratory of Magnetoelectric Physics and Devices (Grant No. 2022B1212010008) and Shenzhen International Quantum Academy. S. X. X. was also supported by Postdoctoral Science Foundation of China (Grant No. 2022M72071).

## Figure Captions:

**FIG. 1. Resistivity and infrared optical spectroscopy.** (a) Temperature-dependent resistivity $\rho(T)$ (red curve) and its derivative $d\rho/dT(T)$ (blue curve) of $La_4Ni_3O_{10}$. Inset shows the crystal structure of $La_4Ni_3O_{10}$. (b) Reflectivity spectra $R(\omega)$ of $La_4Ni_3O_{10}$ below 5000 $cm^{-1}$ at several fixed temperatures with electric field nearly parallel to *ab* plane. The inset shows the reflectivity from 40 to 30000 $cm^{-1}$. (c) Optical conductivity $\sigma_1(\omega)$ of $La_4Ni_3O_{10}$ at different temperatures below 3000 $cm^{-1}$. The inset displays the experimental $\sigma_1(\omega)$ (black curve) at 15 K and the Drude-Lorentz fitting result (red curve). The decomposed Drude and Lorentz components are also displayed. (d) Spectral weight $S(\omega)$ of $La_4Ni_3O_{10}$ below 6000 $cm^{-1}$ obtained by integrating $\sigma_1(\omega)$. The inset shows the ratio of spectral weight before and after phase transition $S$(15 K)/$S$(150 K). (e) The normalized spectral weight at fixed cut-off frequency as a function of temperature. (f) The normalized Drude weight $\omega_p^2$ (gray point) and scatting rate $1/\tau$ (red point) as a function of temperature.

**FIG. 2. Band structure and ratio of $K_{exp}/K_{band}$.** (a) Brillouin zone and (b) band structure for $La_4Ni_3O_{10}$. (c) The calculated (solid lines) and measured (dashed line) optical conductivity $\sigma_1(\omega)$ of $La_4Ni_3O_{10}$. The inset is the enlargement of Fig. 2(c). (d) Ratio of the experimental kinetic energy and theoretical kinetic energy from band theory $K_{exp}/K_{band}$ for $La_4Ni_3O_{10}$ and various other materials. The values of $K_{exp}/K_{band}$ for other materials are obtained from Refs. 15 and 47.

**FIG. 3. Temperature-dependent transient reflectivity variation $\Delta R/R(t)$.** (a) Relative differential reflectivity transients $\Delta R/R(t)$ at different temperatures and fixed pump fluence of 0.2 $\mu J/cm^2$. (b) Contour map of $\Delta R/R$ as a function of temperature and time delay. (c) Typical temporal evolution $\Delta R/R$ in $La_4Ni_3O_{10}$ at selected temperatures. The solid lines are the fitting curves.

**FIG. 4. The evolution of the amplitude and relaxation time**. (a) Amplitude $A_1$ of fast process. (b) Relaxation time $\tau_1$ of fast process. (c) Amplitude $A_2$ of slow process. (d) Relaxation time $\tau_2$ of slow process. The red solid lines in (a) and (b) are the fitting curves according to Rothwarf-Taylor model.

**Table. 1.** The key parameters of Drude-Lorentz fit at different temperatures. $\omega_0$, $\omega_p$

and Γ refer to central frequency, plasma frequency and scattering rate $1/\tau$, respectively.

| | Drude1 | | Drude2 | | Lorentz (SDW) | | |
|---|---|---|---|---|---|---|---|
| T (K) | $\omega_{P1}$ (cm$^{-1}$) | $\Gamma_1$ (cm$^{-1}$) | $\omega_{P2}$ (cm$^{-1}$) | $\Gamma_2$ (cm$^{-1}$) | $\omega_0$ (cm$^{-1}$) | $\omega_P$ (cm$^{-1}$) | Γ (cm$^{-1}$) |
| 15 | 10700 | 141 | 1990 | 386 | 988 | 9690 | 1120 |
| 80 | 11190 | 137 | 2240 | 472 | 932 | 9440 | 1210 |
| 120 | 10230 | 151 | 2750 | 483 | 904 | 9390 | 1240 |
| 150 | 13280 | 213 | 9620 | 1220 | - | - | - |
| 200 | 12090 | 253 | 11360 | 1190 | - | - | - |
| 250 | 12220 | 303 | 11650 | 1190 | - | - | - |
| 300 | 11580 | 303 | 12260 | 1530 | - | - | - |

**Table. 2.** The low-frequency ($\omega$ < 1000 cm$^{-1}$) Lorentz components at different temperatures.

| | Lorentz1 | | | Lorentz2 | | | Lorentz3 (phonon) | | |
|---|---|---|---|---|---|---|---|---|---|
| T (K) | $\omega_{01}$ (cm$^{-1}$) | $\omega_{P1}$ (cm$^{-1}$) | $\Gamma_1$ (cm$^{-1}$) | $\omega_{02}$ (cm$^{-1}$) | $\omega_{P2}$ (cm$^{-1}$) | $\Gamma_2$ (cm$^{-1}$) | $\omega_{02}$ (cm$^{-1}$) | $\omega_{P2}$ (cm$^{-1}$) | $\Gamma_2$ (cm$^{-1}$) |
| 15 | 307 | 2890 | 175 | 620 | 1660 | 86 | 634 | 660 | 16 |
| 80 | 304 | 2890 | 174 | 619 | 2200 | 134 | 634 | 568 | 15 |
| 120 | 343 | 3620 | 243 | 601 | 3450 | 218 | 633 | 456 | 14 |
| 150 | 368 | 1510 | 117 | 600 | 2000 | 148 | 631 | 406 | 11 |
| 200 | 395 | 1650 | 121 | 603 | 2040 | 148 | 631 | 430 | 11 |
| 250 | 399 | 1700 | 106 | 625 | 2500 | 156 | 628 | 260 | 7 |
| 300 | 376 | 3410 | 269 | 638 | 3180 | 251 | 622 | 390 | 12 |

# Figure 1

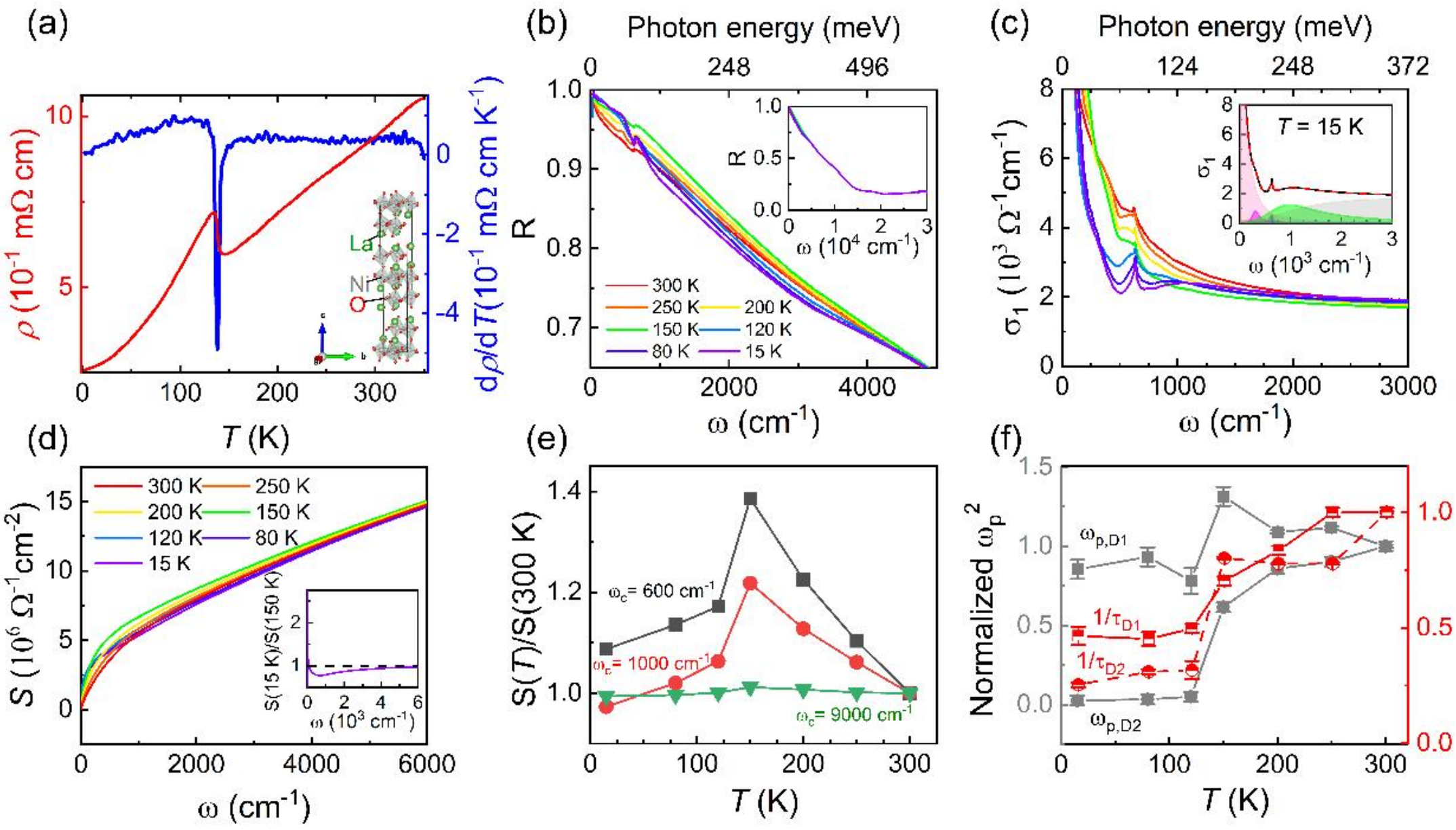
(a)
(b)
(c)
(d)
(e)
(f)
T (K)
ρ (10^-1 mΩ cm)
dρ/dT(10^-1 mΩ cm K^-1)
La
Ni
O
Photon energy (meV)
R
ω (cm^-1)
300 K
250 K
200 K
150 K
120 K
80 K
15 K
σ1 (10^3 Ω^-1cm^-1)
T = 15 K
ω (10^3 cm^-1)
ω (10^4 cm^-1)
S (10^6 Ω^-1cm^-2)
S(15 K)/S(150 K)
S(T)/S(300 K)
ωc= 600 cm^-1
ωc= 1000 cm^-1
ωc= 9000 cm^-1
Normalized ωp^2
Normalized 1/τ
ωp,D1
ωp,D2
1/τD1
1/τD2

# Figure 2

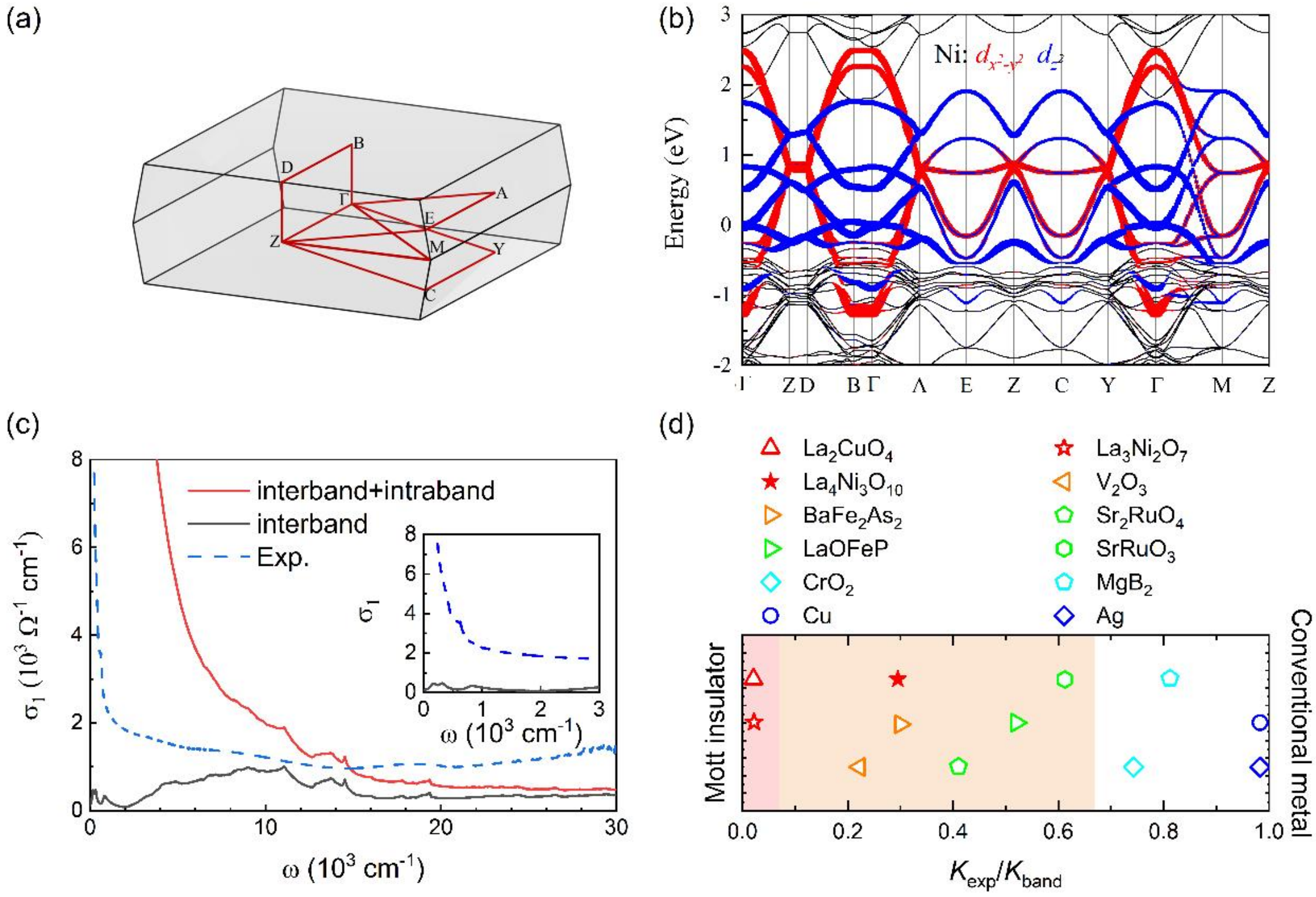

(a)
B
D
Γ
A
E
Z
M
Y
C
(b)
Ni: $d_{x^2-y^2}$ $d_{z^2}$
Energy (eV)
Γ Z D B Γ Λ E Z C Y Γ M Z
(c)
interband+intraband
interband
Exp.
$\sigma_1$ ($10^3$ $\Omega^{-1}$ $cm^{-1}$)
ω ($10^3$ $cm^{-1}$)
$\sigma_1$
ω ($10^3$ $cm^{-1}$)
(d)
$La_2CuO_4$
$La_3Ni_2O_7$
$La_4Ni_3O_{10}$
$V_2O_3$
$BaFe_2As_2$
$Sr_2RuO_4$
LaOFeP
$SrRuO_3$
$CrO_2$
$MgB_2$
Cu
Ag
Mott insulator
Conventional metal
$K_{exp}/K_{band}$

# Figure 3

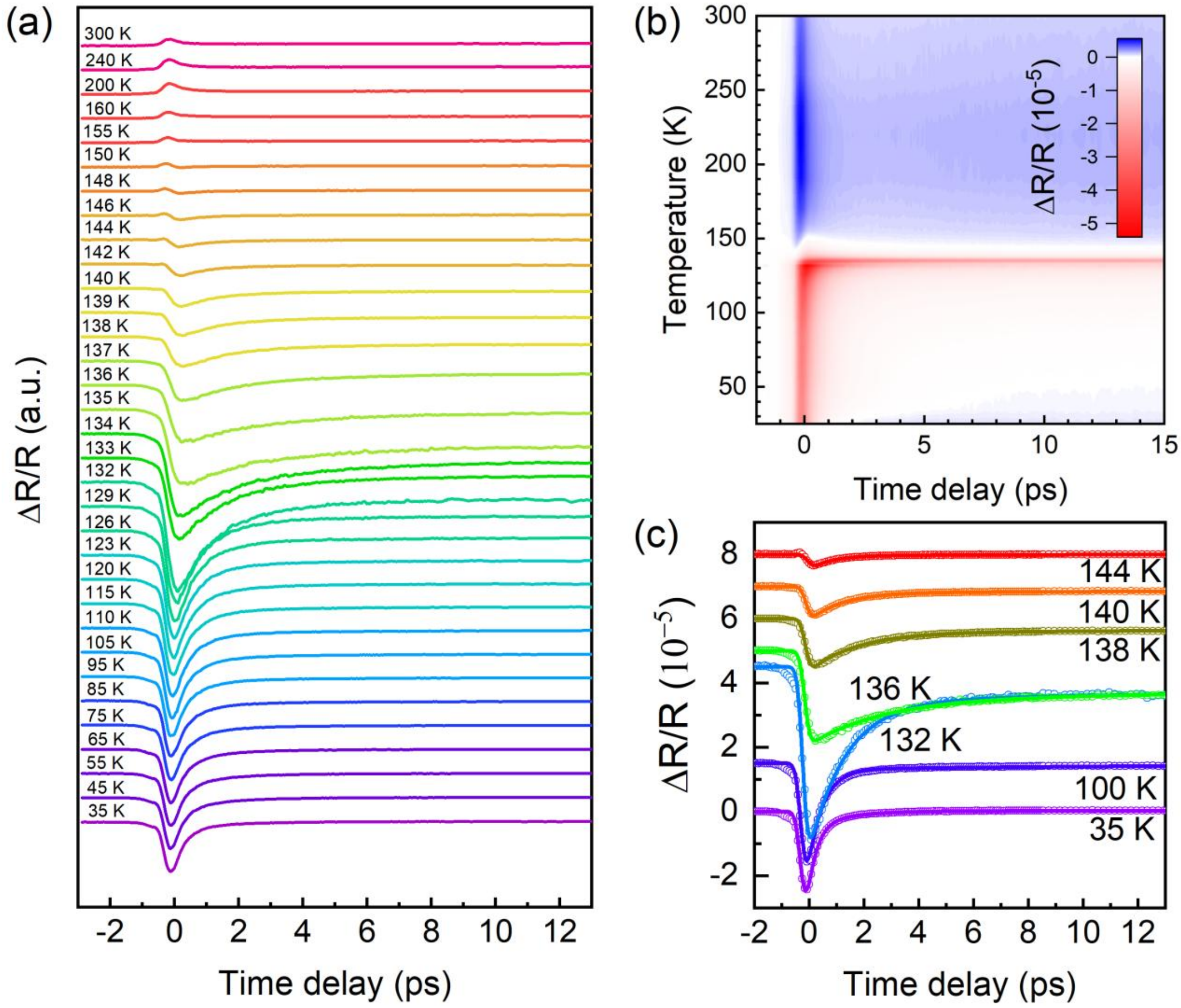
(a)
300 K
240 K
200 K
160 K
155 K
150 K
148 K
146 K
144 K
142 K
140 K
139 K
138 K
137 K
136 K
135 K
134 K
133 K
132 K
129 K
126 K
123 K
120 K
115 K
110 K
105 K
95 K
85 K
75 K
65 K
55 K
45 K
35 K
ΔR/R (a.u.)
-2 0 2 4 6 8 10 12
Time delay (ps)
(b)
Temperature (K)
300 250 200 150 100 50
0 5 10 15
Time delay (ps)
ΔR/R (10^-5)
0 -1 -2 -3 -4 -5
(c)
ΔR/R (10^-5)
8 6 4 2 0 -2
144 K
140 K
138 K
136 K
132 K
100 K
35 K
-2 0 2 4 6 8 10 12
Time delay (ps)

# Figure 4

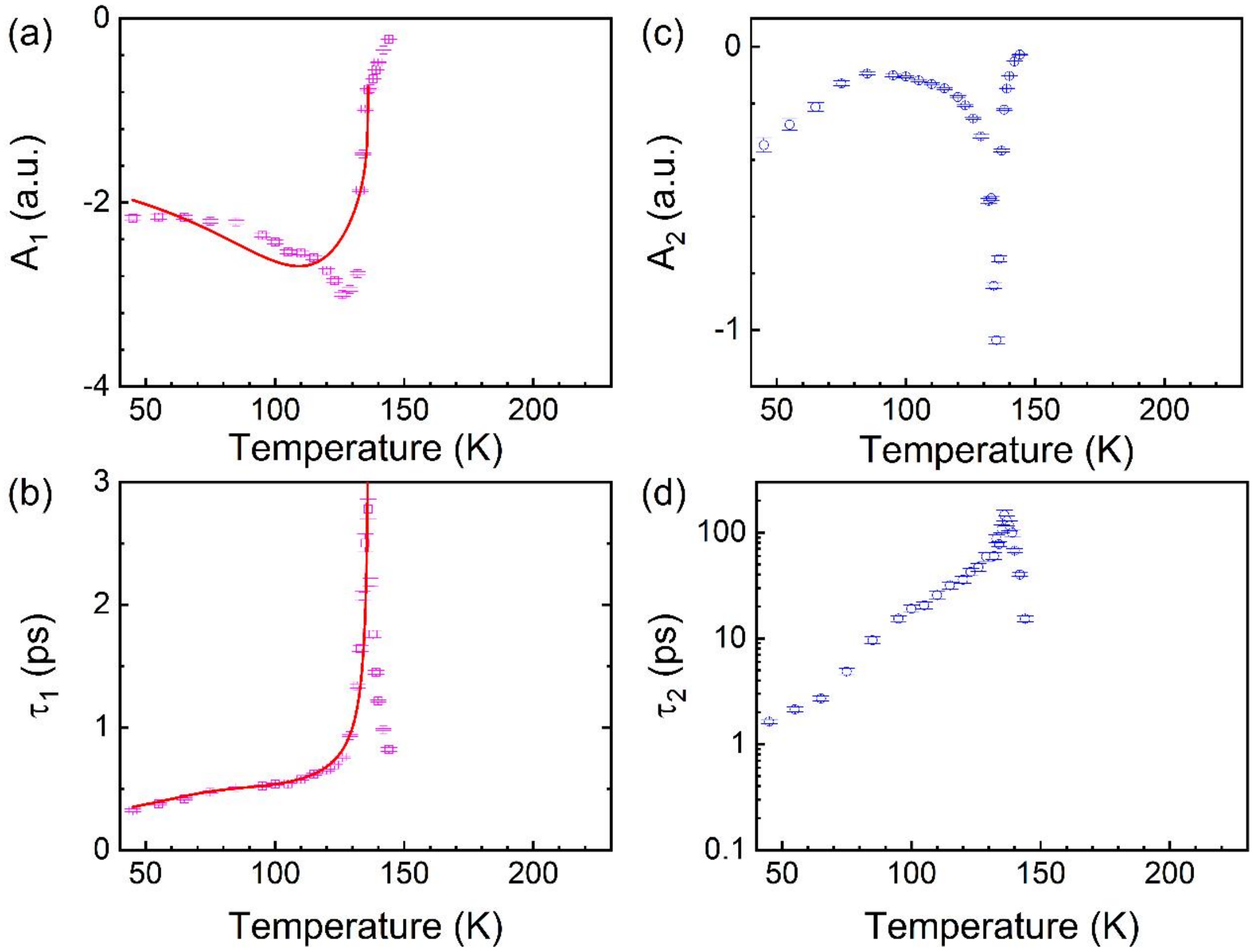

(a)
$A_1$ (a.u.)
0
-2
-4
50
100
150
200
Temperature (K)
(c)
$A_2$ (a.u.)
0
-1
50
100
150
200
Temperature (K)
(b)
$\tau_1$ (ps)
3
2
1
0
50
100
150
200
Temperature (K)
(d)
$\tau_2$ (ps)
100
10
1
0.1
50
100
150
200
Temperature (K)

## Appendix A: the Drude-Lorentz fit

Figure 5 shows the experimental $\sigma_1(\omega)$ and the Drude-Lorentz fitting result. For $La_4Ni_3O_{10}$, there are multiple bands crossing the Fermi energy meaning that one Drude model cannot fit optical conductivity well, thus two Drude peaks are used. After the DW phase transition, one of Drude components decreases sharply.

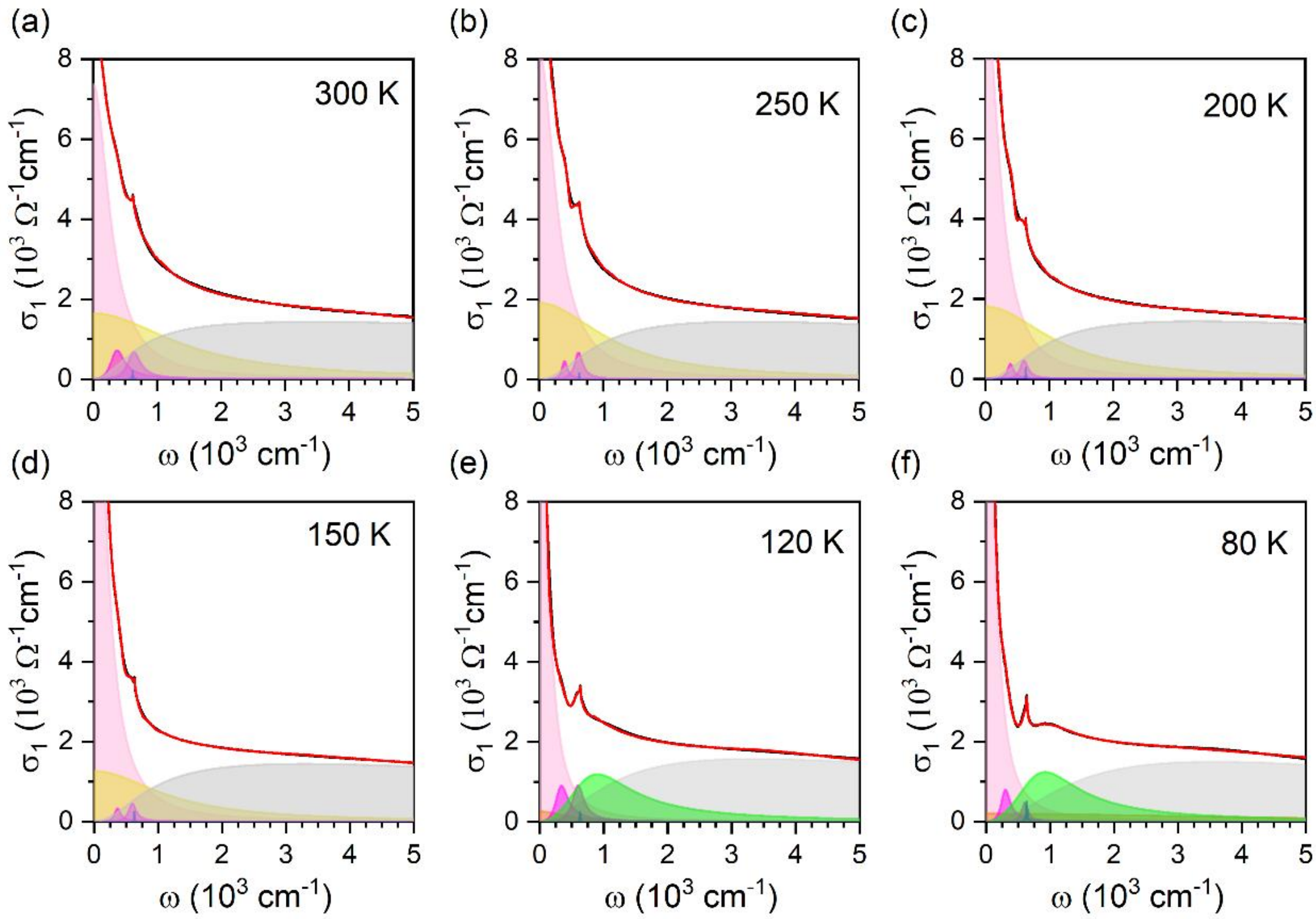

**FIG.5.** The experimental $\sigma_1(\omega)$ (black curve) and the Drude-Lorentz fitting result (red curve). The decomposed Drude and Lorentz components are also displayed: (a) 300 K; (b) 250 K; (c) 200 K; (d) 150 K; (e) 120 K; (f) 80 K, respectively.

## Appendix B: the extended Drude model characteristic

Figure 6 shows the frequency-dependent scattering rate with the extended Drude model. According to the extended Drude model [47], the frequency-dependent scattering rate can be obtained by

$$\frac{1}{\tau(\omega)} = \frac{\omega_p^2}{4\pi}\frac{\sigma_1(\omega)}{|\sigma_1(\omega)|^2}$$

where $\omega_p$ is the plasma frequency, complex conductivity $\sigma = \sigma_1 + i\sigma_2$. $\omega_p$ is calculated by the sum rule. At low frequency below 400 cm$^{-1}$, the scattering rate decreases with cooling and doesn't reach zero at 15 K. After the DW phase transition, a broad peak appears near 1000 cm$^{-1}$ and becomes stronger with cooling which indicates an energy gap possibly opens near the Fermi surface. The extended Drude model is applicable to these correlated materials with only a single energy band crossing the Fermi energy and the decrease in plasma frequency can be considered to be entirely caused by the correlation interaction [48]. For these correlated materials with multiple bands passing through the Fermi energy such as Fe-based superconductors, the scattering rate of different bands is different revealing the extended Drude model with only one scattering rate is meaningless. For $La_4Ni_3O_{10}$, there are multiple bands crossing the Fermi energy meaning that the extended Drude model is not a good approach.

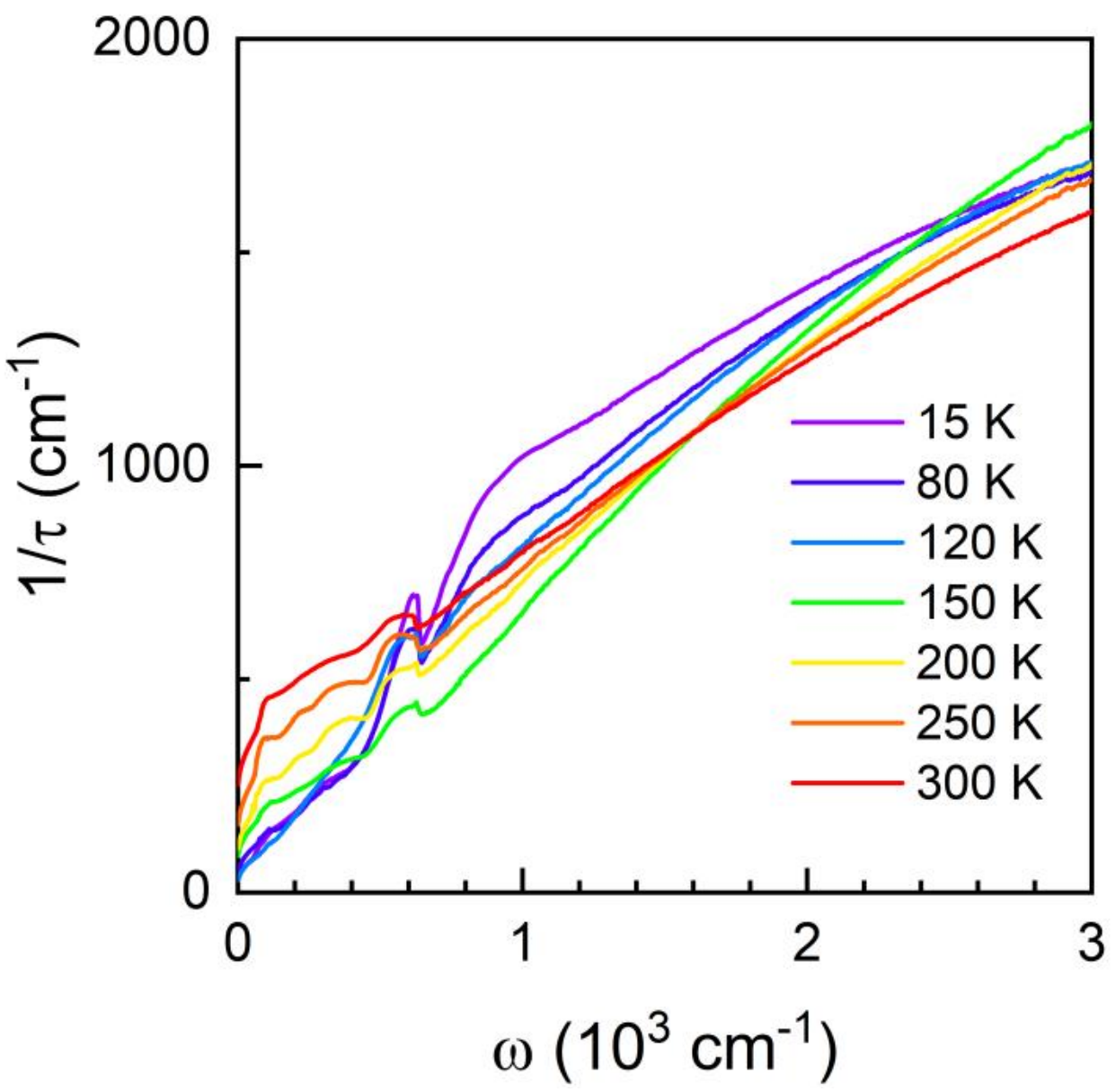

**FIG. 6.** The frequency-dependent scattering rate at different temperatures.